\documentclass[final,5p,times,twocolumn]{elsarticle}
\usepackage{lineno,hyperref,amsmath,amsthm,amssymb,amsfonts,ragged2e,color,subfig}
\usepackage[flushleft]{threeparttable}
\usepackage{booktabs,caption}
\journal{``The European Physical Journal D"}
\begin{document}
\begin{frontmatter}
\title{First and second-order dust-ion-acoustic rogue waves in non-thermal plasma}
\author{S. Banik$^{*1,2}$,  R.K. Shikha$^{**,2}$, A.A. Noman$^{***,2}$, N.A. Chowdhury$^{\dag,3}$, A. Mannan$^{\ddag,4}$,
T.S. Roy$^{\S,5}$, and A.A. Mamun$^{\S\S,2}$}
\address{$^1$Health Physics Division, Atomic Energy Centre, Dhaka-1000, Bangladesh\\
$^2$Department of Physics, Jahangirnagar University, Savar, Dhaka-1342, Bangladesh\\
$^3$Plasma Physics Division, Atomic Energy Centre, Dhaka-1000, Bangladesh\\
$^4$Institut f{\"u}r Mathematik, Martin Luther Universit{\"a}t Halle-Wittenberg, D-06099 Halle, Germany\\
$^5$Department of Physics, Bangladesh University of Textiles, Tejgaon Industrial Area, Dhaka, Bangladesh\\
e-mail: $^*$bsubrata.37@gmail.com, $^{**}$shikha261phy@gmail.com, $^{***}$noman179physics@gmail.com,\\
$^{\dag}$nurealam1743phy@gmail.com, $^{\ddag}$abdulmannan@juniv.edu, $^{\S}$tanu.jabi@gmail.com, $^{\S\S}$mamun\_phys@juniv.edu}
\begin{abstract}
A nonlinear Schr\"{o}dinger equation (NLSE) has been derived by employing reductive
perturbation method for investigating the modulational instability of dust-ion-acoustic waves (DIAWs) in a
four-component plasma having stationary negatively charged dust grains,  inertial warm ions,
and inertialess non-thermal electrons and positrons. It is observed that under consideration, the plasma system
supports both modulationally stable and unstable domains, which are determined by the sign of the dispersive
and nonlinear coefficients of NLSE, of the DIAWs. It is also found that the nonlinearity as well as the height
and width of the first and second-order rogue waves increases with the non-thermality of electron and positron.
The relevancy of our present investigation to the observations in space plasmas is pinpointed.
\end{abstract}
\end{frontmatter}
\section{Introduction}
\label{2:Introduction}
The existence of massive dust grains in different electron-positron-ion (EPI)
plasmas (viz., Jupiter's magnetosphere \cite{Paul2016}, Milky Way \cite{Sardar2016}, auroral zone \cite{Banerjee2016},
accretion disks near neutron stars \cite{Sardar2016}, the hot spots on dust rings in the galactic
centre \cite{Banerjee2016,Paul2017,Jehan2009,Saberian2017}, interstellar medium \cite{Sardar2016},
and around  pulsars \cite{Jehan2009}, etc.) does not only change the dynamics of the plasma medium but
also significantly modifies the basic properties of electrostatic
dust-acoustic (DA) waves (DAWs) \cite{Jehan2009,Saberian2017,Kalejahi2012} and
dust-ion-acoustic (DIA) waves (DIAWs) \cite{Paul2016,Sardar2016,Banerjee2016}. Esfandyari-Kalejahi \textit{et al.} \cite{Kalejahi2012} studied
large amplitude DA solitary waves in EPI plasma, and demonstrated
that the amplitude of the DA solitary waves increases with the charge state of dust grains.
El-Tantawy \textit{et al.} \cite{Tantawy2011} investigated  DIAWs in
EPI dusty plasma medium (EPIDPM), and observed that the amplitude
and width of the positive potential increases with the number density and charge state of the dust grains.

The signature of non-thermal electrons in space plasmas has been observed by the Viking \cite{Bostrom1992}
and Freja satellites \cite{Dovner1994}. Cairns \textit{et al.} \cite{Cairns1995} first introduced non-thermal distribution
and associated parameter $\alpha$ demonstrating the measurement of deviation from Maxwellian distribution
for explaining the high-energy tails in space plasmas. Banerjee and Maitra \cite{Banerjee2016} investigated
DIA solitons and double layers in presence of non-thermal positrons and electrons.
Paul \textit{et al.} \cite{Paul2017} considered a four-component
plasma model having warm ions, stationary dust grains, non-thermal electrons and positrons, and studied DIAWs, and
found that the system supports only positive potential super-solitons.

The modulational instability (MI) of electrostatic waves and formation of
associated rogue waves (RWs) have been governed by the nonlinear Schr\"{o}dinger equation (NLSE).
A number of authors studied the MI of various kind of waves in different plasma
medium \cite{Kourakis2003,Fedele2002,Guo2012,Bains2013,El-Labany2015,El-Tantawy2013}.
Guo \textit{et al.} \cite{Guo2012} investigated
the MI of DIAWs in EPIDPM in presence of non-extensive electrons and positrons.
Bains \textit{et al.} \cite{Bains2013} considered iso-thermal electrons and positrons to
observe the MI condition of DIAWs, and found that the critical wave number decreases with ion
temperature. El-Labany \textit{et al.} \cite{El-Labany2015} studied the MI of DAWs in a three-component
plasma medium having non-thermal plasma species, and found that the height of the DA RWs (DARWs)
increases with the non-thermality of plasma species. El-Tantawy \textit{et al.} \cite{El-Tantawy2013}
examined ion-acoustic (IA) super RWs in a two-component non-thermal
plasma having inertial ions and inerialess electrons, and reported that the nonlinearity as well as
MI growth rate of the IA waves increases with the non-thermality of electrons. To the best of
our knowledge, the effects of non-thermal electrons and positrons, and stationary negatively charged massive
dust grains on the MI of DIAWs and associated DIA RWs (DIARWs) have not yet been investigated.
Therefore, in our present work, we will examine the MI of DIAWs and associated DIARWs in a four-component EPIDPM.

The manuscript is organized in the following order: The basic governing equations are
presented in Section \ref{2:Model Equations}. The derivation of the NLSE is demonstrated in
Section \ref{2:Derivation of the NLSE}. The MI of DIAWs is provided
in Section \ref{2:Instability analysis}. The DIARWs are exhibited in Section \ref{2:Rogue waves}.
Finally, the conclusion is presented in Section \ref{2:Conclusion}.
\section{Model Equations}
\label{2:Model Equations}
We consider a four-component unmagnetized plasma model having inertial warm ions, inertialess non-thermal
electrons and positrons, and stationary negatively charged massive dust grains. At equilibrium, the quasi-neutrality condition can be
expressed as $ n_{e0}+Z_dn_{d0}=n_{p0}+Z_i n_{i0}$, where $n_{e0}$, $n_{d0}$, $n_{p0}$, and $n_{i0}$ are, respectively,
the equilibrium number densities of electrons, dust grains, positrons, and ions. $Z_i$  is the charge state of the positive
ion and $Z_d$ is the number of electrons residing on the dust grains surface. The normalized governing equations can be written in
the following form
\begin{eqnarray}
&&\hspace*{-1.3cm}\frac{\partial n_+}{\partial t}+\frac{\partial}{\partial x}(n_+ u_+)=0,
\label{2eq:1}\\
&&\hspace*{-1.3cm}\frac{\partial u_+}{\partial t} + u_+\frac{\partial u_+}{\partial x}+ \sigma n_+\frac{\partial n_+}{\partial x} =-\frac{\partial\phi}
{\partial x},
\label{2eq:2}\\
&&\hspace*{-1.3cm}\frac{\partial^{2} \phi}{\partial x^{2}}=\mu_e n_e +\mu_d-n_+-(\mu_e+\mu_d-1)n_p,
\label{2eq:3}\
\end{eqnarray}
where $n_+$ is the warm ions number density normalized by it's equilibrium value $n_{+0}$;
$u_+$ is the ion fluid speed normalized by the IA wave speed $C_+=(Z_+k_BT_e/m_+)^{1/2}$
(with $T_e$ being the electron temperature, $m_+$ being the ion mass, and $k_B$ being the Boltzmann
constant); $\phi$ is the electrostatic wave potential normalized by $k_BT_e/e$ (with $e$ being the
magnitude of single electron charge); the time and space variables are normalized by
$\omega_{p+}^{-1}=(m_+/4\pi Z_+^2e^2n_{+0})^{1/2}$ and $\lambda_{D+}=(k_BT_+/4\pi Z_+n_{+0}e^2)^{1/2}$,
respectively, and $T_+$ being the ion temperature; $p_+=p_{+0}(N_+/n_{+0})^\gamma$ [with $p_{+0}$ being
the equilibrium adiabatic pressure of the ion, and $\gamma=(N+2)/N$, where $N$ is the degree of freedom.
For one-dimensional case: $N=1$ then $\gamma=3$, and $p_{+0}=n_{+0}k_BT_+$]. Other plasma parameters are considered as
$\sigma=3T_+/Z_+T_e$, $\mu_e=n_{e0}/Z_+n_{+0}$, and $\mu_d=Z_dn_{d0}/Z_+n_{+0}$.
The expression for the number density of electron (following the Cairns'
non-thermal distribution \cite{Cairns1995,Chowdhury2018a}) can be written as
\begin{eqnarray}
&&\hspace*{-1.3cm}n_e=\Bigg[1-\frac{4\alpha_e}{1+3\alpha_e}\phi+\frac{4\alpha_e}{1+3\alpha_e}\phi^{2}\Bigg]\mbox{exp}(\phi),
\nonumber\\
&&\hspace*{-0.9cm}= 1+F_1\phi+F_2\phi^{2}+F_3\phi^{3}+\cdot\cdot\cdot,
\label{2eq:4}\
\end{eqnarray}
where $F_1=(1-\alpha_e)/(1+3\alpha_e)$, $F_2=1/2$, $F_3=(15\alpha_e+1)/(18\alpha_e+6)$, and $\alpha_e$ being the non-thermality of electrons.
The number density of positron (following the Cairns' non-thermal distribution \cite{Cairns1995,Chowdhury2018a}) can be written as
\begin{eqnarray}
&&\hspace*{-1.3cm}n_p=\Bigg[1+\frac{4\alpha_p}{1+3\alpha_p}\delta\phi+\frac{4\alpha_p}{1+3\alpha_p}\delta^{2}\phi^{2}\Bigg]\mbox{exp}(-\delta\phi),
\nonumber\\
&&\hspace*{-0.8cm}=1+F_4\phi+F_5\phi^{2}+F_6\phi^{3}+\cdot\cdot\cdot,
\label{2eq:5}\
\end{eqnarray}
where $F_4=[(\alpha_p-1)\delta]/(1+3\alpha_p)$, $F_5=\delta^2/2$,
$F_6=[(15\alpha_p+1)\delta^3]/(18\alpha_p+6)$, and $\alpha_p$ being the
non-thermality of positrons. The ratio of $T_e$ to $T_p$ (positron temperature)
is defined by $\delta=T_e/T_p$. By substituting Eq. \eqref{2eq:4} and
\eqref{2eq:5} into Eq. \eqref{2eq:3} and expanding up to third order in $\phi$,
we can write
\begin{eqnarray}
&&\hspace*{-1.3cm}\frac{\partial^{2} \phi}{\partial x^{2}}+n_+=1+\mu_d+F_7\phi+F_8\phi^2+F_9\phi^3+\cdot\cdot\cdot,
\label{2eq:6}\
\end{eqnarray}
where
\begin{eqnarray}
&&\hspace*{-1.3cm}F_8=\frac{\mu_e-\eta\delta^{2}}{2},
\nonumber\\
&&\hspace*{-1.3cm}F_7=\frac{\mu_e(1-\alpha_e)(1+3\alpha_p)-\eta(\alpha_p-1)(1+3\alpha_e)\delta}{(1+3\alpha_e)(1+3\alpha_p)},
\nonumber\\
&&\hspace*{-1.3cm}F_9=\frac{\mu_e(15\alpha_e+1)(18\alpha_p+6)+\eta(15\alpha_p+1)(18\alpha_e+6)\delta^3}{(18\alpha_e+6)(18\alpha_p+6)},
\nonumber\
\end{eqnarray}
where $\eta=\mu_e+\mu_d-1$, and the terms containing $F_7$, $F_8$, and $F_9$ in Eq. \eqref{2eq:6}
are due to the contribution of the non-thermal electrons and positrons.
\section{Derivation of the NLSE}
\label{2:Derivation of the NLSE}
To study the MI of DIAWs, we will derive a standard NLSE by employing the reductive perturbation method. So, we
first introduce the stretched co-ordinates \cite{Chowdhury2018a,Chowdhury2018b}
\begin{eqnarray}
&&\hspace*{-1.3cm}\xi=\epsilon(x-v_gt),
\label{2eq:7}\\
&&\hspace*{-1.3cm}\tau=\epsilon^{2} t,
\label{2eq:8}\
\end{eqnarray}
where $v_g$ is the group speed and $\epsilon$ is a small parameter.
We can write the dependent variables as \cite{C1,C2,C3,C4}
\begin{eqnarray}
&&\hspace*{-1.3cm}n_+=1+\sum_{m=1}^{\infty}\epsilon^{m}\sum_{l=-\infty}^{\infty}n_{+l}^{(m)}(\xi,\tau)\mbox{exp}[i l(kx-\omega t)],
\label{2:eq9}\\
&&\hspace*{-1.3cm}u_+=\sum_{m=1}^{\infty}\epsilon^{m}\sum_{l=-\infty}^{\infty}u_{+l}^{(m)}(\xi,\tau)\mbox{exp}[i l(kx-\omega t)],
\label{2:eq10}\\
&&\hspace*{-1.3cm}\phi=\sum_{m=1}^{\infty}\epsilon^{m}\sum_{l=-\infty}^{\infty}\phi_l^{(m)}(\xi,\tau)\mbox{exp}[i l(kx-\omega t)],
\label{2:eq11}\
\end{eqnarray}
where $k$ ($\omega$) is real variable representing the carrier wave number (frequency).
The derivative operators in the above equations are treated as follows \cite{C5,C6,C7,C8}:
\begin{eqnarray}
&&\hspace*{-1.3cm}\frac{\partial}{\partial x}\rightarrow\frac{\partial}{\partial x}+\epsilon\frac{\partial}{\partial\xi},
\label{2eq:12}\\
&&\hspace*{-1.3cm}\frac{\partial}{\partial t}\rightarrow\frac{\partial}{\partial t}-\epsilon v_g \frac{\partial}
{\partial\xi}+\epsilon^2\frac{\partial}{\partial\tau}.
\label{2eq:13}\
\end{eqnarray}
Now, by substituting Eqs. \eqref{2eq:7}$-$\eqref{2eq:13}, into Eqs. \eqref{2eq:1}, \eqref{2eq:2},
and \eqref{2eq:6}, and selecting the terms containing $\epsilon$, the first order ($m=1$
and $l=1$) reduced equations can provide the dispersion relation of DIAWs
\begin{eqnarray}
&&\hspace*{-1.3cm}\omega^{2}=\frac{k^{2}}{k^{2}+F_7}+\sigma k^{2}.
\label{2eq:14}\
\end{eqnarray}
We again consider the second harmonic with ($m=2$ and $l=1$) and
with the compatibility condition, we have obtained the group velocity of DIAWs
\begin{eqnarray}
&&\hspace*{-1.3cm}v_g=\frac{\omega^{2}-\beta^{2}}{k \omega},
\label{2eq:15}\
\end{eqnarray}
where $\beta=\omega^{2}-\sigma k^{2}$.
The amplitude of the second-order harmonics is found to be proportional to $|\phi_1^{(1)}|^2$
\begin{eqnarray}
\hspace*{-0.5cm}n_{+2}^{(2)}=F_{10}|\phi_1^{(1)}|^2,~~~~~n_{+0}^{(2)}=F_{13}|\phi_1^{(1)}|^2,
\label{2eq:16}\\
\hspace*{-0.5cm}u_{+2}^{(2)}=F_{11}|\phi_1^{(1)}|^2,~~~~~u_{+0}^{(2)}=F_{14}|\phi_1^{(1)}|^2,
\label{2eq:17}\\
\hspace*{-0.5cm}\phi_{2}^{(2)}=F_{12}|\phi_1^{(1)}|^2,~~~~~\phi_{0}^{(2)}=F_{15}|\phi_1^{(1)}|^2,
\label{2eq:18}\
\end{eqnarray}
where
\begin{eqnarray}
&&\hspace*{-1.3cm}F_{10}=\frac{\sigma k^6+3\omega^2 k^4+ 2 F_{12}\beta^{2}k^{2}}{2\beta^3},
\nonumber\\
&&\hspace*{-1.3cm}F_{11}=\frac{\omega F_{10}\beta^{2}-\omega k^{4}}{k\beta^{2}},
\nonumber\\
&&\hspace*{-1.3cm}F_{12}=\frac{k^{4}(3\omega^{2}+ \sigma k^{2})-2 F_8 \beta^{3}}{2\beta^2[(4k^2+F_7)\beta-k^{2}]},
\nonumber\\
&&\hspace*{-1.3cm}F_{13}=\frac{2v_g\omega k^{3}+k^{2}\omega^{2}+\sigma k^{4}+F_{15}\beta^{2}}{(v_g^2-\sigma)\beta^{2}},
\nonumber\\
&&\hspace*{-1.3cm}F_{14}=\frac{v_gF_{13}\beta^{2}-2\omega k^{3}}{\beta^{2}},
\nonumber\\
&&\hspace*{-1.3cm}F_{15}=\frac{k^{2}(2\omega v_g k+\sigma k^{2}+\omega^{2})-2 F_8 (v_g^{2}-\sigma)\beta^{2}}
{\beta^{2} [F_7(v_g^{2}-\sigma)-1]}.
\nonumber\
\end{eqnarray}
Finally, the third harmonic modes ($m=3$) and ($l=1$), with the help of
Eqs. \eqref{2eq:14}$-$\eqref{2eq:18}, give a set of equations, which can be
reduced to the following NLSE:
\begin{eqnarray}
&&\hspace*{-1.3cm}i\frac{\partial\Phi}{\partial\tau}+P\frac{\partial^2\Phi}{\partial\xi^2}+ Q\Phi|\Phi|^2=0,
\label{2eq:19}\
\end{eqnarray}
we have considered $\Phi$=$\phi_1^{(1)}$ for simplicity, and in Eq. \eqref{2eq:19}, $P$ is the dispersion coefficient which can be written as
\begin{eqnarray}
&&\hspace*{-1.3cm}P=\frac{\beta(4\sigma\omega k^{2}-3v_gk\omega^{2}-\sigma v_g k^{3})}{2\omega^{2} k^{2}},
\label{2eq:20}\
\end{eqnarray}
and $Q$ is the nonlinear coefficient which can be written as
\begin{eqnarray}
&&\hspace*{-1.3cm}Q=\frac{\beta^{2}(3 F_9+2 F_8 F_{12}+2 F_8F_{15})-Q^{\prime}}{2\omega k^{2}},
\label{2eq:21}\
\end{eqnarray}
where $Q^{\prime}=k^2[(\sigma k^{2}+\omega^{2})(F_{10}+F_{13})+2\omega k(F_{11} +F_{14})]$.
The space and time evolution of the DIAWs are directly governed by the coefficients $P$ and $Q$.
\begin{figure}[t!]
\centering
\includegraphics[width=80mm]{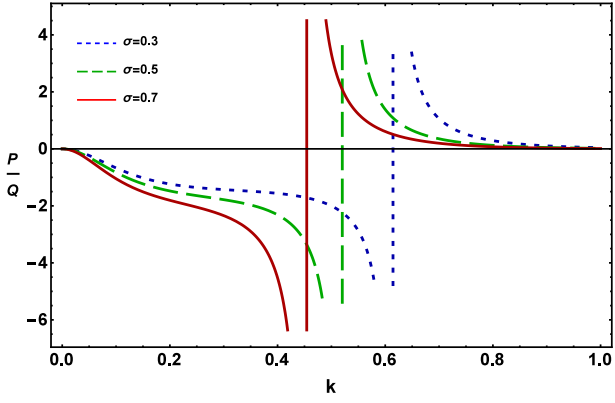}
\caption{The variation of $P/Q$ with $k$ for different values of
$\sigma$ when $\alpha_e=0.5$, $\alpha_p=0.5$, $\delta=1$, $\mu_d=0.02$, and $\mu_e=1.5$.}
\label{2Fig:F1}
\end{figure}
\begin{figure}[t!]
\centering
\includegraphics[width=80mm]{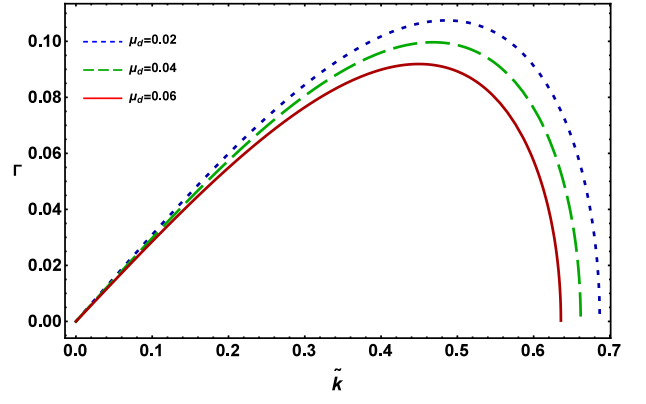}
\caption{The variation of $\Gamma$ with $\widetilde{k}$ for different values of $\mu_d$ when $\alpha_e=0.5$, $\alpha_p=0.5$, $ \delta=1$, $\mu_e=1.5$,
$\sigma=0.3$, $\phi_0=0.5$, and $k=0.7$.}
 \label{2Fig:F2}
\end{figure}
\begin{figure}[t!]
\centering
\includegraphics[width=80mm]{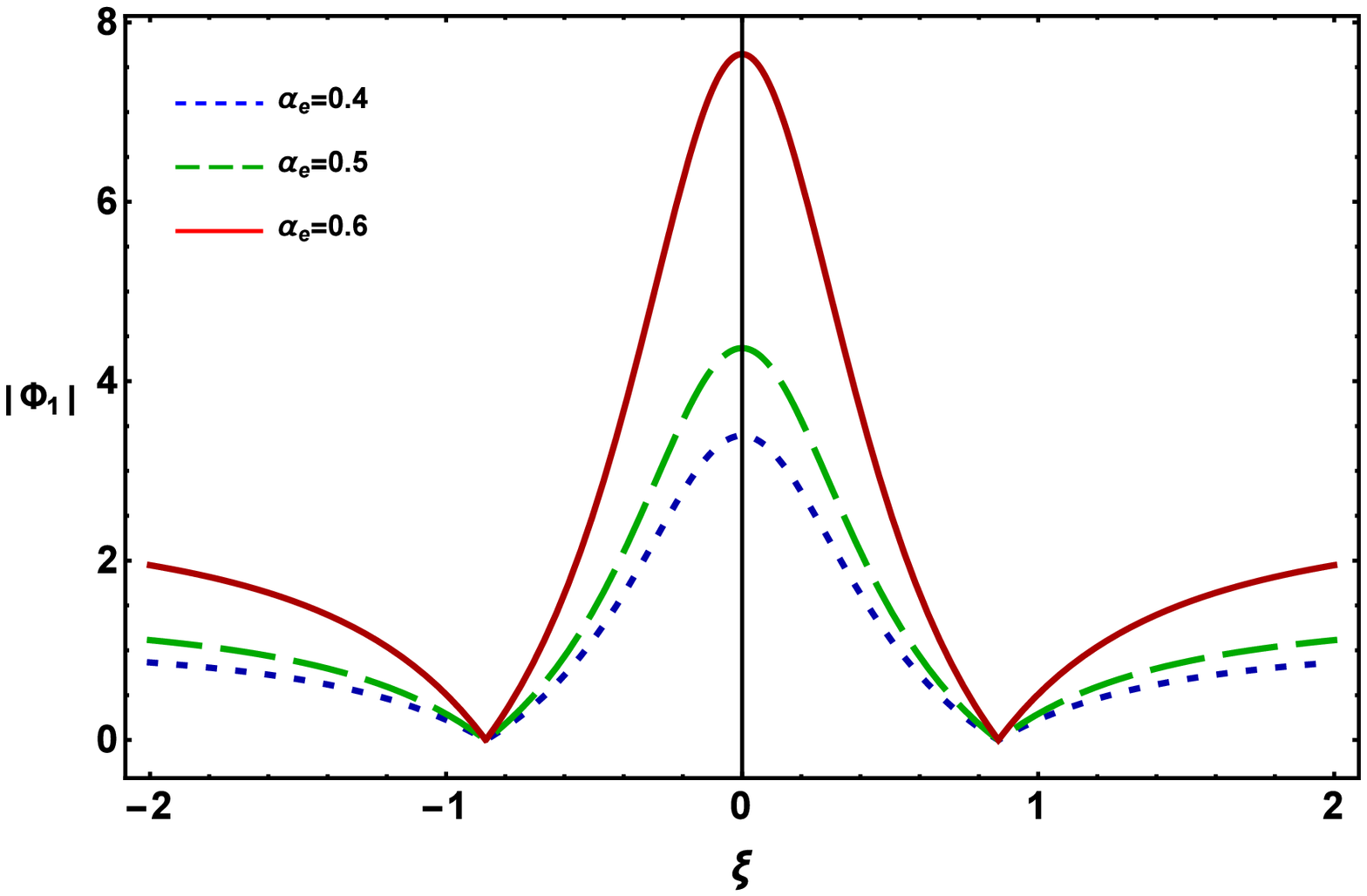}
\caption{The variation of $|\phi_1|$ with $\xi$ for different
values of $\alpha_e$ when  $\alpha_p=0.5$, $\delta=1$, $\mu_d=0.02$, $\mu_e=1.5$, $\sigma=0.3$,
$\tau=0$, and $k=0.7$.}
\label{2Fig:F3}
\end{figure}
\begin{figure}[t!]
\centering
\includegraphics[width=80mm]{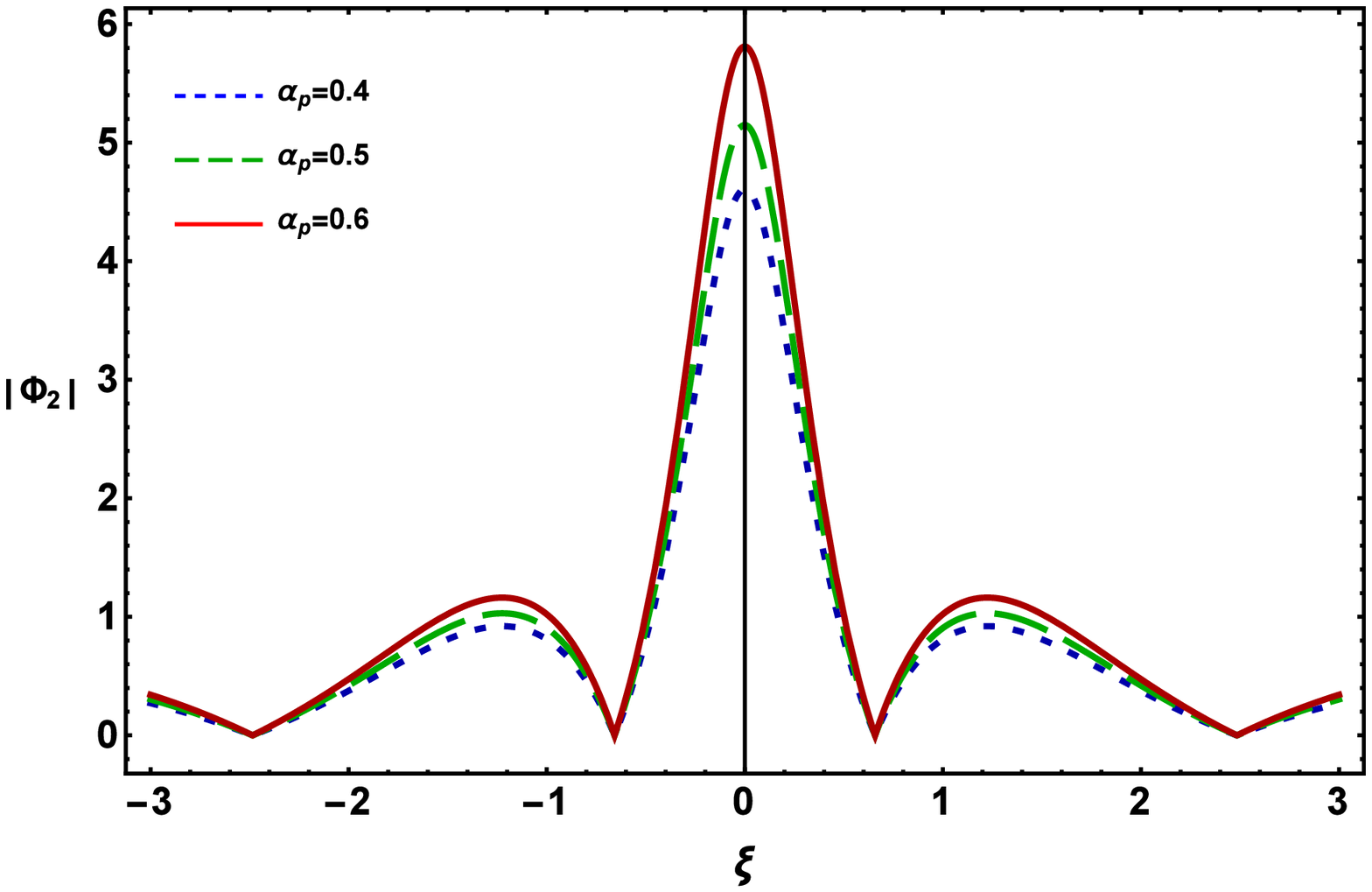}
\caption{The variation of $|\phi_2|$ with $\xi$ for different
values of $\alpha_p$ when  $\alpha_e=0.5$, $\delta=1$, $\mu_d=0.02$, $\mu_e=1.5$, $\sigma=0.3$,
$\tau=0$, and $k=0.7$.}
 \label{2Fig:F4}
\end{figure}
\begin{figure}[t!]
\includegraphics[width=80mm]{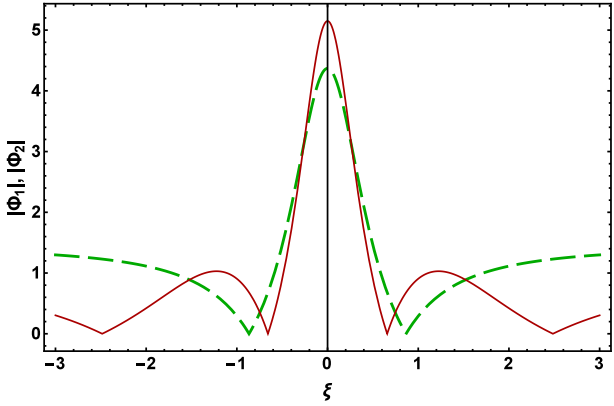}
\caption{The variation of first-order (dashed green curve) and
second-order (solid red curve) rational solutions of NLSE at
$k=0.7$ and $\tau=0$.}
 \label{2Fig:F5}
\end{figure}
\section{Instability analysis}
\label{2:Instability analysis}
To study the MI of DIAWs, we consider the linear solution of the Eq.
\eqref{2eq:19} in the form $\Phi=\widetilde{\Phi}e^{iQ|\widetilde{\Phi}|^2\tau}$+c.c.,
where $\widetilde{\Phi}=\widetilde{\Phi}_0+\epsilon\widetilde{\Phi}_1$
and $\widetilde{\Phi}_1=\widetilde{\Phi}_{1,0}e^{i(\widetilde{k}\xi-\widetilde{\omega}{
\tau})}+c.c$. We note that the amplitude depends on the frequency, and that the perturbed
wave number $\widetilde{k}$ and frequency $\widetilde{\omega}$ which are different from $k$ and
$\omega$. Now, substituting these into Eq. \eqref{2eq:19}, one can easily obtain the following
nonlinear dispersion relation \cite{Chowdhury2018a,Chowdhury2018b}
\begin{eqnarray}
&&\hspace*{-1.3cm}\widetilde{\omega}^{2}=P^{2}\widetilde{k}^{2}\Big(\widetilde{k}^{2}-\frac{2|\widetilde{\Phi}_0|^{2}}{P/Q}\Big).
\label{2eq:22}\
\end{eqnarray}
It is observed here that the ratio $P/Q$ is negative (i.e.,
$P/Q<0$), the DIAWs will be modulationally stable. On the other
hand, if the ratio $P/Q$ is positive (i.e., $P/Q>0$), the DIAWs
will be modulationally unstable. We have graphically examined
the effect of temperature of the ion and electron as well as the charge state of the warm positive ion in recognizing the stable
(i.e., $k<k_c$) and unstable (i.e., $k>k_c$) domains of DIAWs in Fig. \ref{2Fig:F1}, and it is
clear from this figure that (a) the plasma system under consideration supports the DIAWs
with either stable  (i.e., $P/Q<0$) or unstable  (i.e., $P/Q>0$); (b) the stable domain increases with the
increase in the value of the charge state of the warm positive ion when $T_+$ and $T_e$ are invariant;
(c) the $k_c$ decreases with the increase in the value of the ion temperature while increases with $T_e$
for a fixed value of $Z_+$, and this result agrees with the result of Bains \textit{et al.} \cite{Bains2013}.

It is obvious from Eq. \eqref{2eq:22} that the DIAWs becomes modulationally unstable
when $\widetilde{k}_c>\widetilde{k}$ in the regime $P/Q>0$, where
$\widetilde{k}_c = \sqrt{2(Q/P)}{|\widetilde{\Phi}_0|}$. The
growth rate $\Gamma$ of the modulationally unstable DIAWs is \cite{Guo2012,Chowdhury2018a,Chowdhury2018b}
given by
\begin{eqnarray}
&&\hspace*{-1.3cm}\Gamma=|P|\widetilde{k}^{2}\sqrt{\frac{\widetilde{k}_c^{2}}{\widetilde{k}^{2}}-1}.
\label{2eq:23}\
\end{eqnarray}
The variation of the $\Gamma$ with $\widetilde{k}$ for different values of $\mu_d$ can be seen in Fig. \ref{2Fig:F2}. It is obvious from
this figure that (a) the maximum value of $\Gamma$ decreases (increases) with the increase in the values of $Z_d$ ($Z_+$) for a constant value of
$n_{d0}$ and $n_{+0}$; (b) as we increase the value of $n_{d0}$ ($n_{+0}$), the maximum value of the $\Gamma$ decreases (increases)
when $Z_d$ and $Z_+$ remain constant. The physics of this result is that
the nonlinearity of the plasma system increases with the charge state and number density of the
warm ion, but decreases with the charge state and number density of the stationary negatively charged massive dust grains.
\section{Rogue waves}
\label{2:Rogue waves}
The NLSE \eqref{2eq:19} has a variety of rational solutions, among them
there is a hierarchy of rational solution that are localized in
both the $\xi$ and $\tau$ variables.  The first-order rational solution of
Eq. \eqref{2eq:19} can be written as \cite{Ankiewiez2009}
\begin{eqnarray}
&&\hspace*{-1.3cm}\Phi_1(\xi,\tau)= \sqrt{\frac{2P}{Q}} \Big[\frac{4(1+4iP\tau)}{1+16P^{2}\tau^{2}+4\xi^{2}}-1 \Big]\mbox{exp}(2iP\tau).
\label{2eq:24}\
\end{eqnarray}
Equation \eqref{2eq:24} reveals that a significant amount
of DIAWs energy is concentrated into a comparatively small
region in EPIDPM. We have numerically analysed Eq. \eqref{2eq:24} in Fig. \ref{2Fig:F3}
to illustrate the influence of non-thermal
electrons on the formation of DIARWs, and it can be seen from the figure that (i) the height and width of
the DIARWs increase as we increase in the value of non-thermality of the electrons (via $\alpha_e$);
(ii) The physics of this result is that with increasing the
value of $\alpha_e$, the nonlinearity of the plasma system is increasing, which leads to
increase the height and the width of the DIARWs. This result agrees with the result of El-Labany \textit{et al.} \cite{El-Labany2015}.

The interaction of the two or more first-order RWs can generate higher-order RWs
which has a more complicated nonlinear structure. The second-order rational
solution of Eq. \eqref{2eq:19} can be written as \cite{Ankiewiez2009}
\begin{eqnarray}
&&\hspace*{-1.3cm}\Phi_2 (\xi, \tau)=\sqrt{\frac{P}{Q}}\Big[1+\frac{G_2(\xi,\tau)+iM_2(\xi,\tau)}{D_2(\xi,\tau)}\Big]\mbox{exp}(i\tau P),
\label{2eq:25}\
\end{eqnarray}
where
\begin{eqnarray}
&&\hspace*{-1.3cm}G_2(\xi,\tau)=\frac{-\xi^4}{2}-6(P\xi\tau)^2-10(P\tau)^4
\nonumber\\
&&\hspace*{0.2cm}-\frac{3\xi^2}{2}-9(P\tau)^2+\frac{3}{8},
\nonumber\\
&&\hspace*{-1.3cm}M_2(\xi,\tau)=-P\tau\Big[\xi^4+4(P\xi\tau)^2+4(P\tau)^4
\nonumber\\
&&\hspace*{0.2cm}-3\xi^2+2(P\tau)^2-\frac{15}{4}\Big],
\nonumber\\
&&\hspace*{-1.3cm}D_2(\xi,\tau)=\frac{\xi^6}{12}+\frac{\xi^4(P\tau)^2}{2}+\xi^2(P\tau)^4
\nonumber\\
&&\hspace*{0.2cm}+\frac{\xi^4}{8}+\frac{9(P\tau)^4}{2}-\frac{3(P\xi\tau)^2}{2}
\nonumber\\
&&\hspace*{0.2cm}+\frac{9\xi^2}{16}+\frac{33(P\tau)^2}{8}+\frac{3}{32}.
\nonumber\
\end{eqnarray}
Figure \ref{2Fig:F4} represents the second-order DIARWs associated with DIAWs
in the modulationally unstable domain (i.e., $P/Q>0$). The increase in the value of $\alpha_p$
does not only cause to change the height of the DIARWs but also causes to change the width of the
DIARWs. Figure \ref{2Fig:F5} indicates the first-order and second-order solution of the NLSE at
$\tau = 0$, and it is clear from this figure that (a) the second-order
rational solution has double structures compared
with the first-order rational solution; (b) the height of the
second-order rational solution is always greater than the height of the first-order rational solution; (c) the potential
profile of the second-order rational solution becomes more
spiky (i.e., the taller height and narrower width) than the
first-order rational solution; (d) the second (first) order
rational solution has four (two) zeros symmetrically located
on the $\xi$-axis; (e) the second (first) order rational
solution has three (one) local maxima.
\section{Conclusion}
\label{2:Conclusion}
In this study, we have performed a nonlinear analysis of DIAWs in an unmagnetized EPIDPM
having stationary massive nagetively charged dust grains, inertial warm positive ions,
and inertialess non-thermal Cairns' distributed electrons and positrons. The evolution of
DIAWs is governed by the standard NLSE, and the coefficients $P$ and  $Q$ of NLSE can recognize the modulationally stable and
unstable domains of  DIAWs. It is observed that the critical wave number, for which the MI sets in, decreases with ion temperature but
increases with electron temperature. The nonlinearity as well as the height and width of the DIARWs increases with the non-thermality
of electrons and positrons. The limitation of this work is that the gravitational and magnetic fields are not consider.
In future and for better understanding, someone can
investigate the nonlinear propagation in a four-component EPIDPM by considering the gravitational and magnetic
fields. However, these results may be applicable in understanding the conditions of the MI of DIAWs and associated DIARWs in Jupiter's
magnetosphere \cite{Paul2016}, Milky Way \cite{Sardar2016}, auroral zone \cite{Banerjee2016},
accretion disks near neutron stars \cite{Sardar2016}, the hot spots on dust rings in the galactic
centre \cite{Banerjee2016,Paul2017,Jehan2009,Saberian2017}, interstellar medium \cite{Sardar2016},
and around  pulsars \cite{Jehan2009}, etc.

\end{document}